\documentclass[preprint]{elsarticle}
\usepackage{amsfonts}
\usepackage{amsmath}
\usepackage{amssymb}
\usepackage{pifont}
\usepackage{natbib}
\usepackage{geometry}
\usepackage{graphicx}%
\providecommand{\U}[1]{\protect\rule{.1in}{.1in}}

\begin{document}

\title{Decoherence without decoherence\tnoteref{t1}}
\tnotetext[t1]{Thanks to Miles Blencowe, Cliff Burgess, Paul Davies, Jim Hartle, Lee Smolin,
Max Schlosshauer, Dieter Zeh, and Wojciech Zurek for comments on an early
draft, and to Robin\ Blume-Kohout, Amit Hagar, Adrian Kent, Owen Maroney and Jos Uffink
for interesting and enjoyable discussions.}

\author[PI]{Steven Weinstein}
\ead{sweinstein@perimeterinstitute.ca or sw@uwaterloo.ca}
\address[PI]{Perimeter Institute for Theoretical Physics, 31 Caroline St N, Waterloo ON N2L\ 2Y5 Canada}

\begin{keyword}
decoherence \sep classical \sep quasiclassical \sep emergence
\PACS 03.65.Ta \sep 03.65.Yz \sep 05.30.Ch \sep 98.80.Qc
\end{keyword}

\begin{abstract}
It has been claimed that decoherence of open quantum systems explains the
tendency of macroscopic systems to exhibit quasiclassical behavior. We show
that \emph{quasi}classicality is in fact an unremarkable property, characterizing
generic subsystems of environments even in the absence of dynamical
decoherence. It is suggested that decoherence is best regarded
as explaining the persistence of true classicality, rather than the emergence, rather than the
emergence of quasiclassicality.

\end{abstract}

\maketitle

\section{Introduction}

Over the last four decades \cite{Zeh70}, the study of decoherence has begun to
shed light on the effects of the interaction of open quantum systems with
their environments. It has been shown that, for some interesting model
systems, certain pure states, sometimes called \emph{pointer states}%
\ \cite{Zur81}\cite{Zur82}\cite{Sim78} survive interaction with the
environment without losing their coherence or purity. This in turn implies
that superpositions of these states lose coherence in such a way that the
result is an incoherent, improper \emph{mixture }of such states which is
approximately stable. The fact that macroscopic subsystems interacting with an
appropriate environment can be seen to exhibit decoherence in a preferred
basis, along with the fact that the basis in question often corresponds to a
paradigmatically classical observable such as position, has led to claims that
\textquotedblleft the classical structure of phase space emerges from the
quantum Hilbert space in the appropriate limit\textquotedblright%
\ \cite{Zur03}; that \textquotedblleft the appearance of classicality is
therefore grounded in the structure of the physical laws governing the
system-system environment interactions\textquotedblright\cite{Schl07}; and
that \textquotedblleft there are strong signs that the transition [from
quantum to classical] can be understood as something that emerges quite
naturally and inevitably from quantum theory\textquotedblright\cite{Ball08}.
Other, similar claims lie ready to hand \cite{Zur91}\cite{Joos03}. Thus
classicality is supposed not to be endemic to quantum theory, but to emerge
naturally via certain natural interactions when sufficiently macroscopic
objects interact with their environment.

Criticisms of the decoherence program (see e.g. \cite{Sta06}) have to date
focused largely on the fact that the phenomenon in question is only known to
occur for certain model Hamiltonians. If the properties of these Hamiltonians
are not generic, then the possibility of a general explanation of the
emergence of classicality is undermined. The concern of this paper is,
however, orthogonal. Rather than contesting the generality of the models under
consideration, we will focus on a single canonical model -- the central-spin
model -- and show that the features which the central spin attains via the
decoherence process are equally features of the subsystems of its environment,
implying that there is nothing especially interesting about the
quasiclassicality which is supposed to characterize the central spin in the
wake of decoherence. The truly classical states, in the sense considered here,
will be shown to be the pointer states themselves, states which by definition
do \emph{not} undergo decoherence.

\section{Decoherence and classicality}

The process of decoherence works roughly as follows. Consider a subsystem $S$
with (pure) state $\psi^{S}$ interacting with an environment $E$ with state
$\psi^{E}$. If the subsystem is sufficiently macroscopic, and if the
Hamiltonian governing the combined evolution of subsystem and environment is
appropriate, then the environment as a whole acts as a kind of measuring
device, in that the effective state of the environment (given by its reduced
density matrix) will reliably become correlated with certain subsystem
observables. Which properties of the system are \textquotedblleft
measured\textquotedblright\ by the environment -- which observables (if any)
become nontrivially correlated -- will depend on the Hamiltonian, including
the self-Hamiltonians of system and of environment \cite{TW02}\cite{DHMM05}.
Eigenstates of the subsystem observables in question, the pointer states, will
be stable or approximately stable under such measurement-like interactions,
while arbitrary \emph{superpositions} of pointer states will evolve into
improper \emph{mixtures} of those states as a result of the environment's
correlation with the pointer observable. The tendency for the reduced density
matrix of the subsystem to be driven into a small subset of the available
states by the environment is called \emph{einselection}, short for
\emph{environment induced superselection} \cite{Zur82}\cite{Zur03}.
Decoherence, then, refers to the process by which pure states lose their
coherence, and more particularly to a process which favors a particular basis.

What does this all have to do with classicality?\ The driving idea is that
classicality has to do with the stability over time of typical classical
observables such as position, momentum, or energy.\footnote{See \cite{Jans08}
for a careful and clear discussion of several different senses of
classicality\ that might be brought to bear in this context.} Thus if we have
an interaction which picks out position eigenstates of a certain subsystem as
pointer states, then the subsystem may be said behave classically in case it
is in one of the pointer states, since it can be predicted to have a definite
trajectory in space. Indeed, this is quite similar to the \textquotedblleft
criterion of reality\textquotedblright\ stipulated by Einstein, Podolsky and
Rosen \cite{EPR35}, whereby a property is attributable to a system if it can
be said to possess that property with certainty.

Associating the unambiguous possession a particular property with classicality
is unexceptionable, but the claims in the decoherence literature typically
associate a more general kind of classicality, sometimes distinguished as
\textquotedblleft quasiclassicality\textquotedblright\ \cite{Schl07} with a
larger class of subsystems, those described by improper \emph{mixtures} of
pointer states. The idea behind calling such states \textquotedblleft
quasiclassical\textquotedblright\ seems to be that they behave like ensembles
of classical subsystems with respect to the observable of interest.
Interference effects in particular are wiped out. For example, consider a
two-slit interference experiment with electrons. The introduction of an
appropriate environment such as dust or visible light has the effect of
inducing decoherence and destroying the interference pattern, and the
electrons furthermore behave as if they are members of a classical ensemble of
particles, some of which emerge from one slit and some from the other. Thus
the observed behavior of the electrons has affinities with the behavior of
particles described by classical mechanics, and is said to be quasiclassical.
The \textquotedblleft quasi\textquotedblright\ is in place because the mixture
refers, not to actual ensemble, but to a single system, having been obtained
by tracing out the environmental degrees of freedom with which it is entangled.

Perhaps the simplest example of a system which exhibits decoherence to a
preferred basis is the central spin model. \ Here we find that the central
spin, initially in a pure state, evolves into an incoherent mixture of $z$
eigenstates, unless, that is, the initial state is itself a $z$ eigenstate.
What we will show is that an arbitrary state of the spin's environment -- an
arbitrary state of any other spin -- is \emph{also} an incoherent mixture of
$z$ eigenstates. \ Thus what is offered as a distinctive feature of the
central spin turns out to be a generic feature of an arbitrary subsystem.
\ Thus the supposed property of quasiclassicality is better thought of as a
generic feature of quantum states, one which has to do with the fact that on
any reasonable measure, most subsystems are massively entangled with other
subsystems, and thus \textquotedblleft already-decohered\textquotedblright.

\subsection{Example:\ Central spin model}

Consider for example the so-called central spin model \cite{Zur82} in which
one contemplates a system consisting of $N+1$ two-level systems, $N$ of which
are coupled to a central spin $S$ via the Hamiltonian%
\[
\hat{H}=\frac{1}{2}\hat{\sigma}_{z}\otimes\left(  \sum\limits_{i=1}^{N}%
g_{i}\hat{\sigma}_{z}^{(i)}%
\bigotimes_{i' \not= i}%
\hat{I}_{i^{\prime}}\right)
\]
where $\hat{I}_{i}$ is the identity operator for the $i$'th system. (Here
there is no macroscopic/microscopic distinction; rather, the distinctive
dynamical role of the central spin singles it out as special.) An initial pure
state of the form
\begin{equation}
\psi=\alpha\left\vert +z\right\rangle \left\vert E_{0}\right\rangle
+\beta\left\vert -z\right\rangle \left\vert E_{0}\right\rangle
\label{Centralspin initial}%
\end{equation}
will, via the unitary evolution $U(t)=e^{-i\hat{H}t}$ generated by this
Hamiltonian, evolve toward an entangled state $\psi(t)=\alpha\left\vert
+z\right\rangle \left\vert E_{+}(t)\right\rangle +\beta\left\vert
-z\right\rangle \left\vert E_{-}(t)\right\rangle $. After a sufficient amount
of time $t_{d}$ has passed, $\left\langle E_{+}|E_{-}\right\rangle \approx0$,
and the reduced density matrix of the central spin will be well-approximated
by
\begin{equation}
\rho^{S}=\alpha^{2}\left\vert +z\right\rangle \left\langle +z\right\vert
+\beta^{2}\left\vert -z\right\rangle \left\langle -z\right\vert \text{.}
\label{Centralspin final}%
\end{equation}
One can represent this evolution on the Bloch sphere as the evolution of
initially pure states of the central qubit (the surface of the sphere),
evolving, modulo extremely unlikely Poincare-type fluctuations, toward a
narrow ellipse along the $z$ axis:%

\begin{center}
\includegraphics[scale=1.00]{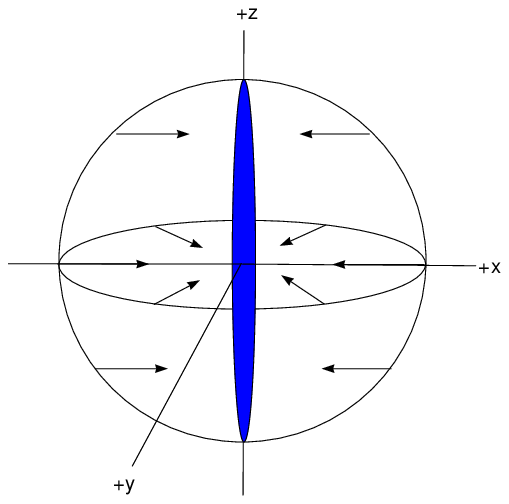}
\\
FIG 1 (color online): Evolution of the central spin. Points away from the z
axis move toward it along $z=$ $constant$ surfaces.
\end{center}

\noindent Though it is no surprise that a subsystem should lose coherence upon
interaction with its environment, and thus move away from the surface of the
Bloch sphere, we have here in addition the phenomenon of einselection, in
which the loss of coherence occurs in a preferred direction. In particular,
the loss of purity is proportional to the angle with the $z$ axis, with the
pointer states $\left\vert +z\right\rangle $ and $\left\vert -z\right\rangle $
suffering no loss whatsoever.

\section{Properties of the environment}

The superselection of a preferred set of states corresponding to eigenstates
or near-eigenstates of typical classical observables is, on the face of it, an
interesting phenomenon suggestive of an emergent quasiclassicality. What we
will now show is that in fact, typical quantum subsystems are in many cases
already in such states -- they are \emph{already decohered}.

Let us proceed by looking at the central spin example in more detail. \ The
initial state of the environment, denoted $\left\vert E_{0}\right\rangle $
above, is a pure state about which we have no information \emph{a priori}.
(Note that, in contrast to our earlier treatment \cite{SW08}, we are granting
the assumption that the system and the environment are initially in a product
state.)\ A \textquotedblleft proper\textquotedblright\ mixture expressing our
ignorance is represented by the density matrix
\[
\Omega_{E}=%
{\displaystyle\sum\limits_{i=1}^{d}}
\frac{1}{d}\left\vert \psi_{i}\right\rangle \left\langle \psi_{i}\right\vert
\]
where $d=2^{n}$ is the dimension of the Hilbert space of the $n$ qubits that
make up the environment, and where the $\left\vert \psi_{i}\right\rangle $ are
orthonormal basis vectors for this space. This corresponds to an unbiased
probability distribution with respect to the unitarily invariant Haar measure,
reflecting our complete ignorance as to which pure state the system is in at
the outset. It is \textquotedblleft maximally-mixed\textquotedblright,
exhibiting random behavior with respect to any choice of observable, and its
von Neumann entropy is therefore maximal.

Let us now inquire as to the description of an arbitrarily chosen subsystem
$e_{1}$ of the environment $E$, where $H_{E}=H_{e_{1}}\otimes...\otimes
H_{e_{n}}$, corresponding to the $n$ spins which make up the environment.
Given the mixture\ $\Omega_{E}$, the effective state of $e_{1}$ is the reduced
state $\Omega_{e_{1}}=Tr_{\tilde{E}}\Omega_{E}$, where $\tilde{E}$ refers to
the rest of the environment ($H_{\tilde{E}}=H_{e_{2}}\otimes...\otimes
H_{e_{n}}$). Since $\Omega_{e}$ is a multiple of the identity, so too is
$\Omega_{e_{1}}$; both are maximally mixed states.

Suppose, now, the environment starts out in some particular unknown pure state
$\rho_{E}$. We can think of this, if we like, as obtained by random sampling
from the distribution $\Omega_{E}$. Thus a particular environmental spin
$e_{1}$will be described by the density matrix $\rho_{e_{1}}=Tr_{\tilde{E}%
}\rho_{E}$. It may be in a pure state, or a mixed state, but it is more likely
to be in a mixed state. \ In fact, it has been shown \cite{Page93}\cite{Pop06}
that the state of the subsystem $\rho_{e_{1}}$ will be almost
indistinguishable from the state $\Omega_{e_{1}}$. More specifically, the
average value of the \textquotedblleft trace distance\textquotedblright%
\ $D(\rho,\Omega):=\frac{1}{2}Tr(\left\vert \rho-\Omega\right\vert )$
\cite{Fuchs95}\cite{NC00} between the two states is bounded by
\begin{equation}
0<\left\langle D(\rho_{e_{1}},\Omega_{\tilde{E}})\right\rangle \leq
\frac{d_{e_{1}}}{2}\sqrt{\frac{1}{d_{\tilde{E}}}}%
\end{equation}
(where $d_{e_{1}}$ and $d_{\tilde{E}}$ are respectively the dimensions of
$e_{1}$ and $\tilde{E}$), so that $D(\rho_{e_{1}},\Omega_{e_{1}})\approx0$ for
almost all states $\rho_{e_{1}}$. In other words, if one takes an arbitrary
pure state of the environment, then an arbitrary small subsystem of the
environment will be very well approximated by a maximally mixed state. \emph{A
fortiori}, in the case of the central spin model, an environmental spin will,
with overwhelming likelihood, live in the superselection sector. \ This is a
purely kinematic fact, involving no dynamics, no loss of quantum coherence.
One might, in the manner of John Wheeler, call this \emph{decoherence without
decoherence}.

It is also salient to note that this feature of subsystems survives the
decohering interaction, for the simple reason that the ensemble from which it
is drawn -- in this case $\Omega_{e_{1}}$ -- remains maximally mixed
throughout the evolution. Thus environmental subsystems will, with
overwhelming probability, reside in the superselection sector which is
supposed to be characteristic of \textquotedblleft classical\textquotedblright%
\ systems, and will remain there indefinitely.

\section{Classicality revisited}

What we have shown is straightforward to the point of being obvious, in
retrospect. \ In the model under consideration, the central spin will evolve
from an initially pure state into a state which resides in the ostensibly
quasiclassical superselection sector, while a random environmental spin will
almost \emph{always} be found in this superselection sector. \ This implies
that \emph{quasi}classicality is not a particularly interesting property.

What \emph{is} interesting, on the other hand, is classicality
\emph{simpliciter}, in which the quantum state assigns the system a definite
value over time. This is of course characteristic of the pointer states, but
\emph{not} of superpositions thereof, and not of environmental spins. What
\textquotedblleft decoherence\textquotedblright\ does to pointer states is in
fact to maintain their classicality by precluding a loss of coherence.
Decoherence does not explain the \emph{emergence }of classicality, but its
\emph{persistence}. It does so by preventing the loss of coherence in the
basis of one or more observables. The emergence of classicality, on the other
hand, appears to await a resolution of the so-called \textquotedblleft
measurement problem\textquotedblright\ -- only when physical properties take
on definite values does one have something resembling a classical world.


\begin{thebibliography}{19}
\providecommand{\natexlab}[1]{#1}
\providecommand{\url}[1]{\texttt{#1}}
\providecommand{\urlprefix}{URL }
\expandafter\ifx\csname urlstyle\endcsname\relax
  \providecommand{\doi}[1]{doi:\discretionary{}{}{}#1}\else
  \providecommand{\doi}[1]{doi:\discretionary{}{}{}\begingroup
  \urlstyle{rm}\url{#1}\endgroup}\fi
\providecommand{\bibinfo}[2]{#2}

\bibitem[{Zeh(1970)}]{Zeh70}
\bibinfo{author}{H.~Zeh}, \bibinfo{title}{On the interpretation of measurement
  in quantum theory}, \bibinfo{journal}{Foundations of Physics}
  \bibinfo{volume}{1} (\bibinfo{year}{1970}) \bibinfo{pages}{69--76}.

\bibitem[{Zurek(1981)}]{Zur81}
\bibinfo{author}{W.~Zurek}, \bibinfo{title}{Pointer basis of quantum apparatus:
  {I}nto what mixture does the wave packet collapse?}, \bibinfo{journal}{Phys.
  Rev.} \bibinfo{volume}{D24} (\bibinfo{year}{1981})
  \bibinfo{pages}{1516--1525}.

\bibitem[{Zurek(1982)}]{Zur82}
\bibinfo{author}{W.~Zurek}, \bibinfo{title}{Environment-induced superselection
  rules}, \bibinfo{journal}{Phys. Rev. D} \bibinfo{volume}{26}
  (\bibinfo{year}{1982}) \bibinfo{pages}{1862}.

\bibitem[{Simonius(1978)}]{Sim78}
\bibinfo{author}{M.~Simonius}, \bibinfo{title}{Spontaneous symmetry breaking
  and blocking of metastable states}, \bibinfo{journal}{Phys. Rev. Lett.}
  \bibinfo{volume}{40} (1978) \bibinfo{pages}{980--983}.

\bibitem[{Zurek(2003)}]{Zur03}
\bibinfo{author}{W.~Zurek}, \bibinfo{title}{{Decoherence, einselection, and the
  quantum origins of the classical}}, \bibinfo{journal}{Rev. Mod. Phys.}
  \bibinfo{volume}{75} (\bibinfo{year}{2003}) \bibinfo{pages}{715}.

\bibitem[{Schlosshauer(2007)}]{Schl07}
\bibinfo{author}{M.~Schlosshauer}, \bibinfo{title}{Decoherence and the
  Quantum-to-Classical Transition}, \bibinfo{publisher}{Springer},
  \bibinfo{address}{Berlin}, \bibinfo{year}{2007}.

\bibitem[{Ball(2008)}]{Ball08}
\bibinfo{author}{P.~Ball}, \bibinfo{title}{Quantum all the way},
  \bibinfo{journal}{Nature} \bibinfo{volume}{453} (\bibinfo{year}{2008})
  \bibinfo{pages}{22}.

\bibitem[{Zurek(1991)}]{Zur91}
\bibinfo{author}{W.~Zurek}, \bibinfo{title}{Quantum measurements and the
  environment-induced transition from quantum to classical}, in:
  \bibinfo{editor}{A.~Ashtekar}, \bibinfo{editor}{J.~Stachel} (Eds.),
  \bibinfo{booktitle}{Conceptual Problems of Quantum Gravity},
  \bibinfo{publisher}{Birkh{\"a}user}, \bibinfo{address}{Boston},
  \bibinfo{pages}{43--66}, \bibinfo{year}{1991}.

\bibitem[{Joos et~al.(2003)Joos, Zeh, Kiefer, Giulini, Kupsch, and
  Stamatescu}]{Joos03}
\bibinfo{author}{E.~Joos}, \bibinfo{author}{H.~Zeh},
  \bibinfo{author}{C.~Kiefer}, \bibinfo{author}{D.~Giulini},
  \bibinfo{author}{J.~Kupsch}, \bibinfo{author}{I.-O. Stamatescu},
  \bibinfo{title}{Decoherence and the Appearance of a Classical World in
  Quantum Theory}, \bibinfo{publisher}{Springer-Verlag},
  \bibinfo{address}{Berlin}, \bibinfo{edition}{2nd} edn., \bibinfo{year}{2003}.

\bibitem[{Stamp(2006)}]{Sta06}
\bibinfo{author}{P.~Stamp}, \bibinfo{title}{The Decoherence Puzzle},
  \bibinfo{journal}{Stud. Hist. Phil. Mod. Phys.} \bibinfo{volume}{37}
  (\bibinfo{year}{2006}) \bibinfo{pages}{467--497}.

\bibitem[{Tessieri and Wilkie(2002)}]{TW02}
\bibinfo{author}{L.~Tessieri}, \bibinfo{author}{J.~Wilkie},
  \bibinfo{title}{Decoherence in a spin--spin-bath model with environmental
  self-interaction} \bibinfo{note}{ArXiv:quant-ph/0209079v1}.

\bibitem[{Dawson et~al.(2005)Dawson, Hines, McKenzie, and Milburn}]{DHMM05}
\bibinfo{author}{C.~Dawson}, \bibinfo{author}{A.~Hines},
  \bibinfo{author}{R.~McKenzie}, \bibinfo{author}{G.~Milburn},
  \bibinfo{title}{Entanglement Sharing and Decoherence in the Spin-Bath},
  \bibinfo{journal}{Phys. Rev. A} \bibinfo{volume}{71} (\bibinfo{year}{2005})
  \bibinfo{pages}{052321}.

\bibitem[{Janssen(2008)}]{Jans08}
\bibinfo{author}{H.~Janssen}, \bibinfo{title}{Reconstructing Reality:
  Environment-Induced Decoherence, the Measurement Problem, and the Emergence
  of Definiteness in Quantum Mechanics} \bibinfo{note}{Draft, available at
  http://philsci-archive.pitt.edu/archive/00004224/}.

\bibitem[{Einstein et~al.(1935)Einstein, Podolsky, and Rosen}]{EPR35}
\bibinfo{author}{A.~Einstein}, \bibinfo{author}{B.~Podolsky},
  \bibinfo{author}{N.~Rosen}, \bibinfo{title}{Can quantum-mechanical
  description of reality be considered complete?}, \bibinfo{journal}{Phys.
  Rev.} \bibinfo{volume}{47} (\bibinfo{year}{1935}) \bibinfo{pages}{777--780}.

\bibitem[{Weinstein(2008)}]{SW08}
\bibinfo{author}{S.~Weinstein}, \bibinfo{title}{Decoherence and the
  (non)emergence of classicality} \bibinfo{note}{ArXiv:0807.3376v2}.

\bibitem[{Page(1993)}]{Page93}
\bibinfo{author}{D.~Page}, \bibinfo{title}{Average Entropy of a Subsystem},
  \bibinfo{journal}{Phys. Rev. Lett.} \bibinfo{volume}{71}
  (\bibinfo{year}{1993}) \bibinfo{pages}{1291--1294}.

\bibitem[{{Popescu} et~al.(2006){Popescu}, {Short}, and {Winter}}]{Pop06}
\bibinfo{author}{S.~{Popescu}}, \bibinfo{author}{A.~J. {Short}},
  \bibinfo{author}{A.~{Winter}}, \bibinfo{title}{{Entanglement and the
  foundations of statistical mechanics}}, \bibinfo{journal}{Nature Physics}
  \bibinfo{volume}{2} (\bibinfo{year}{2006}) \bibinfo{pages}{754--758}.

\bibitem[{Fuchs(1995)}]{Fuchs95}
\bibinfo{author}{C.~Fuchs}, \bibinfo{title}{{Distinguishability and Accessible
  Information in Quantum Theory}} .

\bibitem[{Nielsen and Chuang(2000)}]{NC00}
\bibinfo{author}{M.~Nielsen}, \bibinfo{author}{I.~Chuang},
  \bibinfo{title}{Quantum Computation and Quantum Information},
  \bibinfo{publisher}{Cambridge University Press},
  \bibinfo{address}{Cambridge}, \bibinfo{year}{2000}.

\end{thebibliography}
\end{document}